\documentclass[aps,twocolumn]{revtex4}
\usepackage{epsfig}
\usepackage{times}
\usepackage{hyperref}
\usepackage{color}

\begin{document}

\title{Evolution of the most common English words and phrases over the centuries}

\author{Matja{\v z} Perc}
\thanks{Electronic address: \href{mailto:matjaz.perc@uni-mb.si}{\textcolor{blue}{matjaz.perc@uni-mb.si}}}
\affiliation{Faculty of Natural Sciences and Mathematics, University of Maribor, Koro{\v s}ka cesta 160, SI-2000 Maribor, Slovenia}

\begin{abstract}
By determining which were the most common English words and phrases since the beginning of the 16th century, we obtain a unique large-scale view of the evolution of written text. We find that the most common words and phrases in any given year had a much shorter popularity lifespan in the 16th than they had in the 20th century. By measuring how their usage propagated across the years, we show that for the past two centuries the process has been governed by linear preferential attachment. Along with the steady growth of the English lexicon, this provides an empirical explanation for the ubiquity of the Zipf's law in language statistics and confirms that writing, although undoubtedly an expression of art and skill, is not immune to the same influences of self-organization that are known to regulate processes as diverse as the making of new friends and World Wide Web growth.
\end{abstract}

\maketitle

\section*{1. Introduction}
The evolution of language \cite{nowak_pnas99, hauser_s02, nowak_n02, abrams_n03, sole_n05, lieberman_n07, loreto_np07} is, much like the evolution of cooperation \cite{sigmund_10, nowak_11}, something that distinguishes humans markedly from other species \cite{miller_81, hrdy_11}. While the successful evolution of cooperation enables us to harvest the benefits of collective efforts on an unprecedented scale, the evolution of language, along with the set of grammatical rules \cite{nowak_s01} that allows infinitely many comprehensible formulations \cite{chomsky_65, hauser_96, lightfoot_99, niyogi_06}, enables us to uphold a cumulative culture \cite{lehman_47}. Were it not for books, periodicals and other publications, we would hardly be able to continuously elaborate over what is handed over by previous generations, and consequently, the diversity and efficiency of our products would be much lower than it is today. Indeed, it seems like the importance of the written word for where we stand today as a species cannot be overstated.

The availability of vast amounts of digitised data, also referred to as ``metaknowledge'' or ``big data'' \cite{evans_s11}, along with the recent advances in the theory and modelling of social systems in the broadest possible sense \cite{lazer_s09, castellano_rmp09}, enables quantitative explorations of the human culture that were unimaginable even a decade ago. From human mobility patterns \cite{gonzales_n08, song_s10}, crashes in financial markets \cite{preis_pw11} and our economic life \cite{preis_pta12, preis_sr12}, the spread of infectious diseases \cite{liljeros_mi03, balcan_pnas09, meloni_pnas09} and malware \cite{hu_pnas09, wang_s09}, the dynamics of online popularity \cite{ratkiewicz_prl10} and social movements \cite{holt_pone11}, to scientific correspondence \cite{barabasi_n05, malmgren_s09}, there appear to be no limits to insightful explorations that lift the veil on how we as humans behave, interact, communicate and shape our very existence.

\begin{figure}
\centerline{\epsfig{file=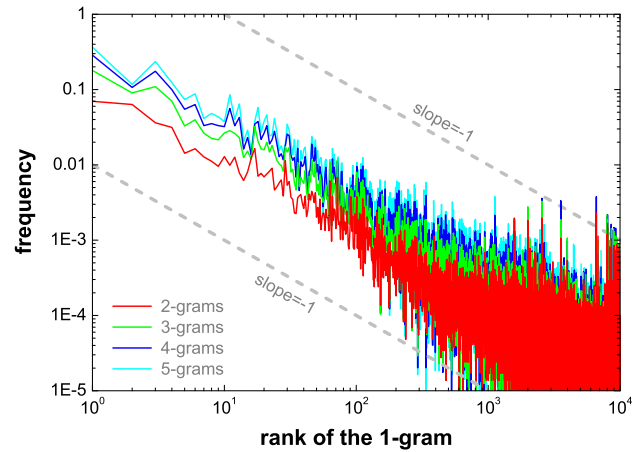,width=8.281cm}}
\caption{Confirmation of the Zipf's law in the examined corpus. By measuring the frequency of $1$-grams in the $n$-grams, where $n>2$ (see figure legend), we find that it is inversely proportional to the rank of the $1$-grams. For all $n$ the depicted curves decay with a slope of $-1$ on a double log scale over several orders of magnitude, thus confirming the validity of the Zipf's law in the examined dataset.}
\label{zipf}
\end{figure}

\begin{figure*}
\centerline{\epsfig{file=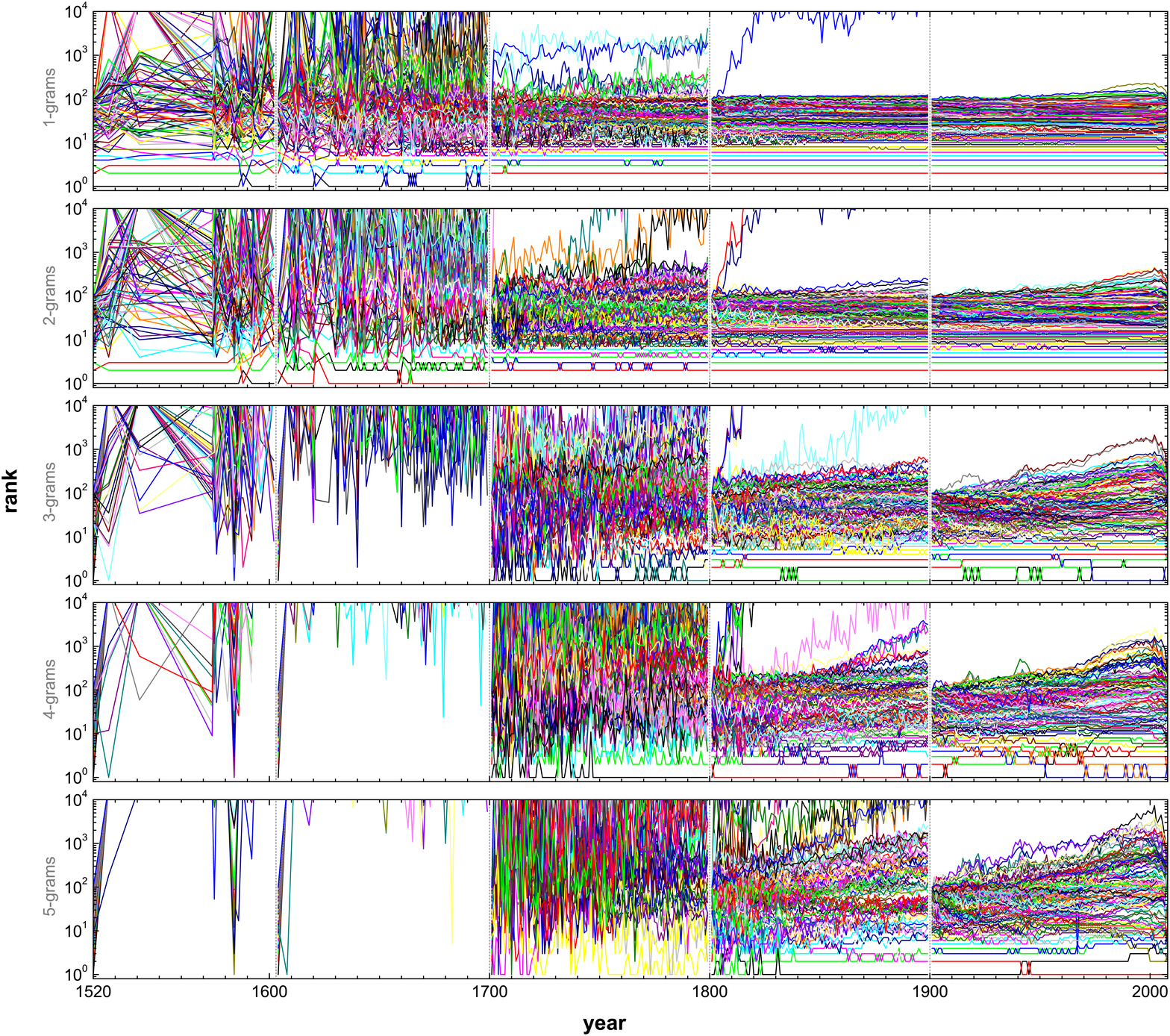,width=16.5cm}}
\caption{Evolution of popularity of the top $100$ $n$-grams over the past five centuries. For each of the five starting years, being $1520$, $1604$, $1700$, $1800$ and $1900$ from left to right (separated by dashed gray lines), the rank of the top $100$ $n$-grams was followed until it exceeded $10.000$ or until the end of the century. From top to bottom the panels depict results for different $n$, as indicated vertically. The advent of the 19th century marks a turning point after which the rankings begun to gain markedly on consistency. Regardless of which century is considered, the higher the $n$ the more fleeting the popularity. Tables listing the top $n$-grams for all available years are available at \href{http://www.matjazperc.com/ngrams}{\textcolor{blue}{http://www.matjazperc.com/ngrams}}.}
\label{popular}
\end{figure*}

Much of what we have learned from these studies strongly supports the fact that universal laws of organization govern how nature, as well as we as a society, work \cite{bak_1996, newman_06}. Languages, as comprehensively reviewed in \cite{sole_jrsi10}, and as suggested already by Zipf \cite{zipf_49} as well as by others before him \cite{manning_99}, are certainly no exception. In fact, in many ways it seem more like it is the other way around. The Zipf's law is frequently related to the occurrence of power-law distributions in empirical data \cite{clauset_siam09}, with examples ranging from income rankings and population counts of cities to avalanche and forest-fire sizes \cite{newman_cp05}. Yet the mechanisms that may lead to the emergence of scaling in various systems differ. The proposal made by Zipf was that there is tension between the efforts of the speaker and the listener, and it has been shown that this may indeed explain the origins of scaling in the human language \cite{cancho_pnas03}. The model proposed by Yule \cite{yule_ptrsb25}, relying on the rich-get-richer phenomenon (see \cite{simkin_pr11} for a review), is also frequently cited as the reason for the emergence of the Zipf's law. With the advent of contemporary network science \cite{watts_n98, barabasi_s99, albert_rmp01}, however, growth and preferential attachment, used ingeniously by Barab{\'a}si and Albert \cite{barabasi_s99} to explain the emergence of scaling in random networks, has received overwhelming attention, also in relation to the emergence of the Zipf's law in different corpora of the natural language \cite{dorogovtsev_prsb01, sole_c10}.

Here we make use of the data that accompanied the seminal study by Michel et al. \cite{michel_s11}, and show empirically, based on a large-scale statistical analysis of the evolution of the usage of the most common words and phrases in the corpus of the English books over the past five centuries, that growth and preferential attachment played a central role in determining the longevity of popularity as well as the emergence of scaling in the examined corpus. The presented results support previous theoretical studies \cite{sole_jrsi10} and indicate that writing, on a large scale, is subject to the same fundamental laws of organization that determine so many other aspects of our existence.

\section*{2. Results}

\begin{figure}
\centerline{\epsfig{file=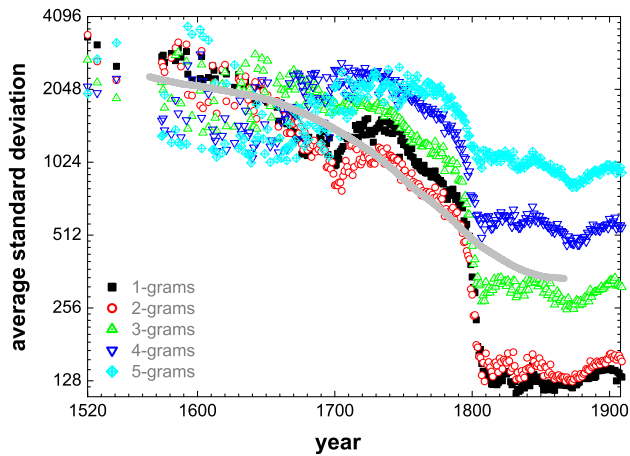,width=8.32cm}}
\caption{``Statistical'' coming of age of the English language. Symbols depict results for different $n$ (see figure legend), as obtained by calculating the average standard deviation of the rank for the top $1000$ $n$-grams 100 years into the future. The thick gray line is a moving average over all the $n$-grams and over the analysis going $50$ and $100$ years into the future as well as backwards. There is a sharp transition to a greater maturity of the rankings taking place at around the year $1800$. Although the moving average softens the transition, it confirm that the ``statistical'' coming of age was taking place and that the 19th century was crucial in this respect.}
\label{mature}
\end{figure}

Henceforth we will, for practical reasons, refer to the words and phrases as $n$-grams \cite{michel_s11}, with the meaning as described in the Methods section. We begin with presenting the results of a direct test of the Zipf's law for the overall most common $1$-grams in the English corpus since the beginning of the 16th century. For this purpose, we treat the $n$-grams for different $n>1$ as individual corpora where the frequencies of the $1$-grams are to be determined. Results presented in Fig.~\ref{zipf} confirm that, irrespective of $n$, the frequency of any given $1$-gram is roughly inversely proportional to its rank. The ragged outlay of the curves is a consequence of the rather special construction of the corpora on which this test was performed. Yet given the time span the data covers as well as its extent, this is surely a very satisfiable outcome of a test for a century old law \cite{manning_99, cancho_acs02} on such a large scale, validating the dataset against the hallmark statistical property of the human language.

Turning to the evolution of popularity, we show in Fig.~\ref{popular} how the rank of the top $100$ $n$-grams, as determined in the years $1520$, $1604$, $1700$, $1800$ and $1900$, varied until the beginning of the next century. During the 16th and the 17th century popularity was very fleeting. Phrases that were used most frequently in $1520$, for example, only intermittently succeeded in re-entering the charts in the later years, despite the fact that we have kept track of the top $10.000$ $n$-grams and have started with the top $100$ $n$-grams in each of the considering starting years. It was not before the end of the 18th century that the top $100$ $n$-grams gradually began succeeding in transferring their start-up ranks over to the next century. The longevity and persistency of popularity is the highest during the 20th century, which is also the last one for which data is available, apart from the eight years into the 21st century. Comparing the different $n$-grams with one another, we find that the $1$-grams were always, regardless of the century considered, more likely to retain their top rankings than the $3$-grams, which in turn outperformed the $5$-grams. This, however, is an expected result, given that single words and short phrases are obviously more likely to be reused than phrases consisting of three, four or even five words.

\begin{figure}
\centerline{\epsfig{file=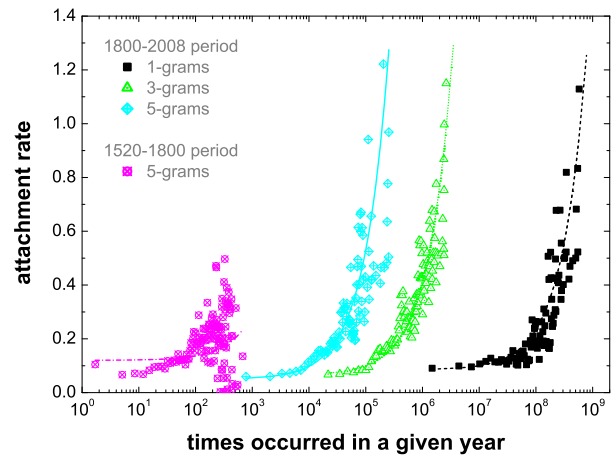,width=8.095cm}}
\caption{Emergence of linear preferential attachment during the past two centuries. Based on the preceding evolution of popularity, two time periods were considered separately, as indicated in the figure legend. While preferential attachment appears to have been in place already during the $1520-1800$ period, large deviations from the linear dependence (the goodness-of-fit is $\approx 0.05$) hint towards inconsistencies that may have resulted in heavily fluctuated rankings. The same analysis for the 19th and the 20th century provides much more conclusive results. For all $n$ the data fall nicely onto straight lines (the goodness-of-fit is $\approx 0.8$), thus indicating that continuous growth and linear preferential attachment have shaped the large-scale organization of the writing of English books over the past two centuries. Results for those $n$-grams that are not depicted are qualitatively identical for both periods of time.}
\label{attach}
\end{figure}

Although the fleeting nature of the top rankings recorded in the 16th and the 17th century is to a degree surely a consequence of the relatively sparse data (only a few books per year) if compared to the 19th and the 20the century, it nevertheless appears intriguing as it is based on the relative yearly usage frequencies of the $n$-grams. Thus, at least a ``statistical'' coming of age of the written word imposes as a viable interpretation. To quantify it accurately, we have performed the same analysis as presented in Fig.~\ref{popular} for the top $1000$ $n$-grams for all years with data, and subsequently calculating the average standard deviation of the resulting $1000$ curves for each starting year. Symbols presented in Fig.~\ref{mature} depict the results of this analysis separately for all the $n$-grams. A sharp transition towards a higher consistency of the rankings occurs at the brink of the 19th century for all $n$, thus giving results presented in Fig.~\ref{popular} a more accurate quantitative frame. These results remain valid if the rankings are traced only $50$ years into the future, as well as if performing the same analysis backwards in time, as evidenced by the thick gray line depicting a moving average over this four scenarios as well as over all the $n$.

Both the validity of the Zipf's law across all the data considered in this study, as well as the peculiar evolution of popularity of the most frequently used $n$-grams over the past five centuries, hint towards large-scale organization gradually emerging in the writing of the English books. Since the groundbreaking work by Barab{\'a}si and Albert on the emergence of scaling in random networks \cite{barabasi_s99}, growth and preferential attachment has become synonymous for the emergence of power laws and leadership in complex systems. Here we adopt this beautiful perspective and test whether it holds true also for the number of occurrences of the most common words and phrases in the English books that were published in the past five centuries. In the seminal paper introducing culturomics \cite{michel_s11}, it was pointed out that the size of the English lexicon has grown by $33 \%$ during the 20th century alone. As for preferential attachment, we present in Fig.~\ref{attach} evidence indicating that the higher the number of occurrences of any given $n$-gram, the higher the probability that it will occur even more frequently in the future. More precisely, for the past two centuries the points quantifying the attachment rate follow a linear dependence, thus confirming that both growth as well as linear preferential attachment are indeed the two processes governing the large-scale organization of writing. Performing the same analysis for the preceding three centuries fails to deliver the same conclusion, although the seed for what will eventually emerge as linear preferential attachment is clearly inferable.

\section*{3. Discussion}
Alone the question ``Which are the most common words and phrases of the English language?'' has a certain appeal, especially if one is able to use digitised data from millions of books dating as far back as the early 16th century \cite{michel_s11} to answer it. On the other hand, writing about the evolution of a language without considering grammar or syntax \cite{chomsky_65}, or even without being sure that all the considered words and phrases actually have a meaning, may appear prohibitive to many outside of the physics community. Yet it is precisely this detachment from detail and the sheer scale of the analysis that enables the observation of universal laws that govern the large-scale organization of the written word. This does not mean that the presented results are no longer valid if we made sure to analyse only words and phrases that actually have meaning or if we had distinguished between capitalized words, but rather that such details don't play a decisive role in our analysis. Regardless of whether a word is an adjective or a noun, or whether it is currently trendy or not, with the years passing by the mechanism of preferential attachment will make sure that the word will obtain its rightful place in the overall rankings. Together with the continuous growth of the English lexicon, we have a blueprint for the emergence of the Zipf's law that is derived from a vast amount of empirical data and supported by theory \cite{barabasi_s99}. This does not diminish the relevance of the tension between the efforts of the speaker and the listener \cite{cancho_pnas03}, but adds to the importance of the analysis of ``big data'' with methods of statistical physics \cite{loreto_jsm11, petersen_sr12} and network science \cite{cancho_prsb01, dorogovtsev_prsb01, sole_c10} for our understanding of the large-scale dynamics of human language.

The allure of universal laws that might describe the workings of our society is large \cite{bak_1996}. Observing the Zipf's law \cite{zipf_49}, or more generally a power-law distribution \cite{newman_cp05}, in a dataset is an indication that some form of large-scale self-organization might be taking place in the examined system. Implying that initial advantages are often self-amplifying and tend to snowball over time, preferential attachment, known also as the rich-get-richer phenomenon \cite{yule_ptrsb25}, the ``Matthew effect'' \cite{merton_sci68}, or the cumulative advantage \cite{price_sci65}, has been confirmed empirically by the accumulation of citations \cite{redner_pt05} and scientific collaborators \cite{jeong_epl03, newman_pnas04}, by the growth of the World Wide Web \cite{newman_06}, and by the longevity of one's career \cite{petersen_pnas11}. Examples based solely on theoretical arguments, however, are many more and much easier to come by. Empirical validations of preferential attachment require large amounts of data with time stamps included. It is the increasing availability of such datasets that appears to fuel progress in fields ranging from cell biology to software design \cite{barabasi_np12}, and as this study shows, helps reveal why the overall rankings of the most common English words and phrases are unlikely to change in the near future, as well as why the Zipf's law emerges in written text.

\section*{Appendix A. Methods}
\subsection*{Raw data}
The seminal study by Michel et al. \cite{michel_s11} was accompanied by the release of a vast amount of data comprised of metrics derived from $\sim4\%$ of books ever published. Raw data, along with usage instructions, is available at \href{http://books.google.com/ngrams/datasets}{\textcolor{blue}{http://books.google.com/ngrams/datasets}} as counts of $n$-grams that appeared in a given corpus of books published in each year. An $n$-gram is made up of a series of $n$ $1$-grams, and a $1$-gram is a string of characters uninterrupted by a space. Although we have excluded $1$-grams that are obviously not words (for example if containing characters outside the range of the ASCII table) from the analysis, some (mostly typos) might have nevertheless found their way into the top rankings. The latter were composed by recursively scanning all the files from the English corpus associated with a given $n$ in the search for those $n$-grams that had the highest usage frequencies in any given year. Tables listing the top $100$, top $1000$ and top $10.000$ $n$-grams for all available years since $1520$ inclusive, along with their yearly usage frequencies and direct links to the Google Books Ngram Viewer, are available at \href{http://www.matjazperc.com/ngrams}{\textcolor{blue}{http://www.matjazperc.com/ngrams}}.

\subsection*{Zipf's law}
Taking the top $10.000$ $n$-grams for all available years as the basis, we have determined the number of unique $n$-grams through the centuries and ranked them according to the total number of occurrences in the whole corpus during all the years. In this way, we have obtained a list of $148.557$ unique $1$-grams, $291.661$ unique $2$-grams, $482.503$ unique $3$-grams, $742.636$ unique $4$-grams, and $979.225$ unique $5$-grams. This dataset was used for testing the Zipf's law by searching for the overall top ranked $1$-grams in all the other $n$-grams ($n>1$) and recording their frequency of occurrence. For example, the $1$-gram ``the'' appeared in $22.826$ of the $291.661$ $2$-grams, hence its frequency is $\sim7.8\%$. By plotting the so obtained frequency in dependence on the rank of the $1$-grams for $n=2,3,4,5$ on a double log scale (see Fig.~\ref{zipf}), we observe four inversely proportional curves, thus confirming the Zipf's law in the constructed dataset.

\subsection*{Attachment rate}
Based on the assumption that the more frequently a given $n$-gram appears the more linked it is to other $n$-grams, we have determined the attachment rate following network science \cite{jeong_epl03} as follows. If an $n$-gram has appeared $m$ times in the year $y$, and $k$ times in the year $y+\Delta y$, the attachment rate is $\alpha(m)=\frac{k}{m \Delta y}$. Note that the occurrences in the data set are not cumulative. Hence there is no difference between $k$ and $m$ in the numerator. Moreover, by the determination of the attachment rate, we are not interested in the relative yearly usage frequencies, but rather in the absolute number of times a given $n$-gram has appeared in the corpus in any given year. Thus $m$ and $k$ are not normalized with the total word counts per year. We have determined $\alpha(m)$ based on the propagation of top $100$ $n$-grams between $1520-1800$ and $1800-2008$ with a yearly resolution. Missing years were bridged by adjusting $\Delta y$ accordingly. For the final display of the attachment rate in Fig.~\ref{attach} and the linear fitting, we have averaged $\alpha(m)$ over $\sim200$ non-overlapping segments in $m$.

\section*{Acknowledgments}
This research was supported by the Slovenian Research Agency (Grant J1-4055).

\end{document}